\documentclass[conference]{IEEEtran}
\IEEEoverridecommandlockouts

\usepackage{cite}
\usepackage{amsmath,amssymb,amsfonts}
\usepackage{algorithmic}
\usepackage{graphicx}
\usepackage{textcomp}
\usepackage{xcolor}
\def\BibTeX{{\rm B\kern-.05em{\sc i\kern-.025em b}\kern-.08em
    T\kern-.1667em\lower.7ex\hbox{E}\kern-.125emX}}
\begin{document}

\title{A Graph Theoretic Approach to Analyze the Developing Metaverse\\
}

\author{\IEEEauthorblockN{Anirudh Dash}
    \IEEEauthorblockA{\textit{Department of Electrical Engineering} \\
        \textit{Indian Institute of Technology, Hyderabad}\\
        Hyderabad, India \\
        ee21btech11002@iith.ac.in}
}

\maketitle

\begin{abstract}
    Despite staggering growth over the past couple of decades, the concept of the metaverse is still in its early stages. Eventually, it is expected to become a common medium connecting every individual. Considering the complexity of this plausible scenario at hand, there's a need to define an advanced metaverse- a metaverse in which, at every point in space and time, two distinct paradigms exist: that of the user in the physical world and that of its real-time digital replica in the virtual one, that can engage seamlessly with each other. The developing metaverse can be thus defined as the transitional period from the current state to, possibly, the advanced metaverse. This paper seeks to model, from a graphical standpoint, some of the structures in the current metaverse and ones that might be key to the developing and advanced metaverses under one umbrella, unlike existing approaches that treat different aspects of the metaverse in isolation. This integration allows for the accurate representation of cross-domain interactions, leading to optimized resource allocation, enhanced user engagement, and improved content distribution. This work demonstrates the usefulness of such an approach in capturing these correlations, providing a powerful tool for the analysis and future development of the metaverse.
\end{abstract}


\section{Introduction}
The metaverse is a persistent, shared, 3D virtual universe that merges with the physical world and offers an immersive space for a diverse range of activities \cite{b1}. Despite rapid growth over the past 20 years, the metaverse remains in its infancy on a global scale. Several features and future possibilities in the realm of the metaverse, as discussed over a decade ago \cite{b7}, have already been implemented, indicating its vast potential. Recent analyses (\cite{b8}, for example) have also examined prospects in this domain.

Recent technical advancements have enabled the use of head-mounted displays (HMDs) while navigating public environments. These devices combine virtual reality (VR) and augmented reality (AR) and provide users with an extended reality (XR). The advanced metaverse can be defined as an XR metaverse where the users interact with the real and virtual worlds simultaneously through avatars (virtual selves of the user). This is a densely populated model, where almost everyone is connected via the virtual world, and their avatar is sensitive to stimuli such as sights and sounds. The actions of individuals in the physical world are reflected in the virtual world.

In the advanced metaverse, HMDs would be used to look at avatars who are present and can interact at various locations in the physical world. Any information that needs to be conveyed to a user can be routed through the avatar. Even inanimate objects will have their virtual counterparts. The above system draws inspiration from Neal Stephenson's science fiction book, Snow Crash. As of today, we are far from such a utopia. From hereon, this paper seeks to model some of the essential aspects that might help us transition from the current state of the metaverse to the advanced metaverse via a graph theoretic approach. Analyzing the metaverse setup graphically and holistically allows one to incorporate key correlations missed in many recent state-of-the-art techniques applied to analogous models within and outside the metaverse.

\section{The Model}

The metaverse can be modeled as a dynamic, temporal (the system evolves as users interact and environments change), multi-layered network, $M(t)=(G_1(t), G_2(t), \cdots G_k(t))$ where each layer $G_i(t)$ represents a unique set of interactions between entities in the space or some form of relationship between them at any given instant of time. Let the graphs in the layers be $G_i(t)(V_i(t), E_i(t))$ where $V_i(t)$ is the set of vertices (users, avatars, objects and environments) of the $i^{th}$ layer and $E_i(t)$ is the set of edges (directed) associated with the nodes in the $i^{th}$ layer, denoting the interactions and relationships (which define edge properties) between various components at timestamp $t$. Note that the vertices can be shared between layers, thus allowing the same vertex to belong to multiple layers. Define the vertex set as $V(t):= \bigcup_{i} V_i(t)$, such that every vertex $v \in V(t)$. Edges can be intra-layer $( u,v \in V_i(t), e_i(t)=(u, v))$ or inter-layer $(u \in V_i(t),v \in V_j(t), e_{ij}(t)=(u_i, v_j))$. From here forth, $(t)$ has been dropped for notational simplicity.



\section{Architectures}

\subsection{Infrastructure}
The infrastructure is critical for supporting the complex and dynamic environment of the advanced metaverse. It encompasses the foundational technologies, platforms, and systems that enable seamless operation, scalability, and performance.

\subsubsection{Communication and Networking}

The layer $G_N(V_N, E_N)$ represents the communication network, where $V_N$ represents networking nodes: servers, routers, and users' devices. The edges represent communication links. The key metrics involved in analyzing these vertices are normalized degree centrality,
betweenness centrality
and clustering coefficients.
Nodes with a high degree centrality are key to network connectivity, while identifying bridges using the betweenness centrality is important for efficient data routing. Clustering coefficients indicate how likely two nodes connected to a common node are also connected to each other.
Optimization of the above network ensures low-latency communication. This requires shortest-path algorithms, dealing with multigraphs and real-time data. Thus, the refined version of Dijkstra's algorithm \cite{b5} can be implemented. Furthermore, load-balancing algorithms (weighted response time, resource-based \cite{b14}) that distribute network traffic evenly across servers and reduce congestion can also be implemented. Flow and cut algorithms, along with minimum spanning trees (MSTs), help identify points of congestion. Presently, queuing models are also used in handling network traffic.

\subsubsection{Distributed and Parallel Computing}

The layer $G_C(V_C, E_C)$ represents the computational resources in the network, where $V_C$ represents cloud instances and specialized servers. For near-optimal resource management, graph partitioning algorithms such as spectral clustering \cite{b16} are employed to divide the graph into subgraphs, whose individual computational needs are easier to manage than the entire system. Algorithms like MapReduce \cite{b30} can be represented as directed acyclic graphs (DAGs), where nodes are tasks and edges are dependencies.  MSTs help establish a backbone network for heavy computation demand nodes. Each node is connected via the minimum total connection cost. This approach ensures high throughput and fault tolerance. Redundant edges can be added to the MST for backup, enhancing reliability. Efficient MST algorithms for temporal graphs already exist \cite{b6}.

\subsubsection{Data Storage and Management}

Graph databases like Neo4j can efficiently manage the relationships between entities in the metaverse. This allows for efficient querying and traversal. The layer $G_S(V_S, E_S)$ represents the storage resources. $V_S$ represents storage nodes such as cloud storage and data centers, while the edges in $E_S$ represent data transfer links. Replica graphs are used to ensure quick access and prevent data loss in the system. Using techniques like eventual consistency in distributed databases, modeled using weakly connected components in graphs, ensures data remains consistent across the network over time. Fast efficient algorithms \cite{b18}, BFS (Breadth First Search), and graph coloring-based algorithms help identify such components.

\subsubsection{Security and Privacy}

Graph-based access control can be used to ensure the security of the networks to which the users and their avatars are connected. Graph-based machine learning algorithms or general graph anomaly detection algorithms \cite{b20} are also used to detect unusual patterns of behavior that may indicate security breaches. For encryption and authentication, public key infrastructure is represented as a layer $G_K(V_K, E_K)$, where the vertices are entities with public keys and edges are trust relationships.

\subsection{Content}

The content layer in the metaverse is a critical component that encompasses all the digital assets, interactions, and experiences that users engage with.

\subsubsection{3D Models and Environments}

Hypergraphs represent polygonal surfaces in 3D models, which are represented as meshes. \cite{b21} demonstrates 3D object recognition via hypergraph analysis. Rendering a 3D model involves the conversion of a mesh into a 2D image. This is optimized by implementing algorithms such as breadth and depth-first searches. Furthermore, to model physical interactions such as movements, several points are taken on the object under consideration and mapped to graphs. Then, its change in connectivity is analyzed to get the requisite results. Hence, this is effectively a variant of the dynamic forest problem. \cite{b22} provides parallel connectivity algorithms for the analysis of such large dynamic graphs.

\subsubsection{User Generated Content}

User Generated Content (UGC) is modeled as a complete social graph (used to model social networks). When users create content, it is often in collaboration with other users or involves the interaction of the user with various other facets in the space. Each collaborative project is a subgraph of the complete graph. DAGs are used for keeping track of different versions of the above content. The edges represent the updates between versions. To share the content, content delivery networks and peer to peer networks are used. Community-finding algorithms are used to foster engagement and collaboration.

\subsubsection{Content Management Systems and Content Delivery Networks}

Content Management Systems are modeled as bipartite graphs, $G_{Con}(V_{Adm} \cup V_{Con}, E_{Con})$, where $V_{Adm}$ are the administrators, $V_{Con}$ are content items, and the edges represent management actions. Access control graphs and workflow graphs are used here. Efficient graph traversal algorithms manage and retrieve content, ensuring users can quickly find and interact with relevant items. Moreover, graph databases store and query metadata, making them relevant and easy to search for.

Content delivery networks (CDNs) involve providing low-latency content to users. They use graph-based caching techniques to ensure that frequently accessed content is at points in the network such that retrieval is fast and efficient. Moreover, there are algorithms \cite{b24} that ensure subgraph-based distribution to minimize latency.

\begin{figure}[t]
    \centering
    \includegraphics[width=0.8\linewidth]{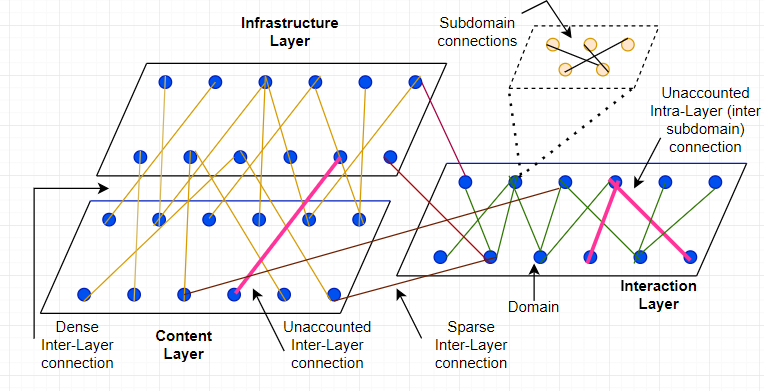}
    \caption{Representation of connections between multiple layers and domains. When the inter-domain/inter-layer connections are higher, missing out on accounting for such connections may lead to incorrect optima.}
    \label{fig:2}
\end{figure}

\subsection{Interaction}

The interaction layer of the metaverse is where users engage with the virtual environment and each other. This layer is closely related to the functioning of the HMDs and how the user can utilize them.

\subsubsection{User Interfaces}

User interfaces can be modeled as graphs $G_{UI}(V_{UI}, E_{UI})$, where the vertices represent interface elements associated with the HMDs, such as buttons, menus, and widgets, while the edges represent actions connecting these vertices- cause and effect- clicks, selections, drags and so on. The graphs used here are mainly state transition graphs. Modifying the graph structure allows for customization and personalization of the user interface.

\subsubsection{Interaction Models and Communication Protocols}

Interaction models ($G_{I}(V_{I}, E_{I})$) are represented using multi-agent graphs where the vertices are the avatars and the edges are the interactions between them. Game theory is used to model and analyze interactions between avatars. Communication protocols involve the use of network and information flow graphs.

\subsubsection{Input Output Systems}

Input and output systems can be represented as a bipartite graph $G_{IO}(V_{in} \cup V_{out}, E_{io})$ where $V_{in}$ represents input devices such as sensors while $V_{out}$ represents output devices such as displays and speakers. The edges represent the data flow between them. Not only this, specific graphs catering solely to the sensor networks ensure comprehensive and accurate data acquisition. Topological graphs can be used for input-output analysis.

\subsubsection{Synchronization Mechanisms}

Ensuring all users in an environment see the same state and experience interactions simultaneously is critical. Graphs (used to maintain synchronization) consist of vertices representing the connected entities, and the edges are data consistency paths. Consensus algorithms \cite{b28} ensure all entities in a distributed system agree on a common state, which helps in real-time interaction.

\subsubsection{Use of Artificial Intelligence}
AI-based behaviors involve integrating artificial intelligence algorithms into virtual entities to mimic human-like actions and decision-making. Decision trees and state machines help in this process. Graph neural networks can assist in several dynamic feature prediction aspects of the metaverse. Reinforcement learning can enable virtual entities to adapt based on metaverse interactions. Natural Language Processing and other behavioral models, such as Bayesian networks or Markov models \cite{b39}, can be critical in keeping such a model going.

\begin{figure}[t]
    \centering
    \includegraphics[width=0.9\linewidth]{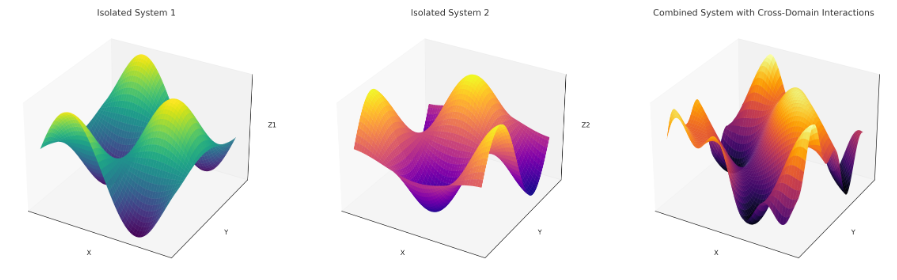}
    \caption{Depiction of variation in cost function when dealing with optimization of isolated systems compared to when all inter-domain interactions are handled.}
    \label{fig:3}
\end{figure}

\section{Advantages}

Consider \cite{b36}, which analyzes UGC and social interactions and their impact on the metaverse. However, it doesn't involve analyzing what kind of data CDNs cache and send. It also doesn't consider "Click Through Rates" \cite{b38} based on the content, which certainly affects user-user interaction. Most of these isolated systems use graph-based techniques, showing the usefulness of graphical models, but fail to capture interdependencies between different domains of the metaverse. \cite{b38} handles multiple subdomains within a larger domain of user interaction, but this doesn't consider the aforementioned factors. Here, $E_{IC}$, $E_{IU}$, $E_{IU}$ associated with $V_I \times V_{con}$, $V_I \times V_{UI}$, $V_{UI} \times V_{con}$ are missing: the graphical techniques suggested, such as centrality measures, would help identify demands that include the effect of these parameters, which might have otherwise been missed and would overload the system, causing it to fail. In the advanced metaverse, a unified model would be key to smooth functioning since user-avatar and avatar-avatar interactions would simultaneously depend on all of the above factors. Moreover, infrastructure hasn't been considered in all of the above. \cite{b39} entails some of the infrastructural requirements to make a simpler version of the above combination work. A unified graphical model encompasses all such factors under one roof, and the optimization thence performed is for a combined cost function, which includes all possibilities, granted the optimization itself may be challenging.

Consider \cite{b35}, which involves a cross-domain optimization in a software-defined Internet of Things. The graphical method is a somewhat analogous, more generalized version for the advanced metaverse. Post some abstraction, their optimization problem is \(\max_{\mathbf{R}_k} \sum_{k=1}^{K} U_k(\mathbf{R}_k)\), where $\mathbf{R}_k$ represents the resource allocation vector within domain $k$ and $U_k(\mathbf{R}_k)$ is the utility function for domain $k$. (\( U_k(\mathbf{R}_k) = {1}/({1 + \exp(-\gamma_k (\mathbf{R}_k - \lambda_k))}) \) where \(\gamma_k\) and \(\lambda_k\) are parameters defining the steepness and midpoint of the utility function for domain \(k\)). It has some constraints: \( \sum_{j} f_j(\mathbf{R}_j) \leq C_l \) where \(f_j(\mathbf{R}_j)\) is the flow of data in domain \(j\) over link \(l\), and \(C_l\) is the capacity of link \(l\). It has an Energy Consumption Model:
\( E_n = \sum_{l \in \text{Links}} \left( e_{tx}(d_{ln}) f_{ln}(\mathbf{R}_l) + e_{rx}(d_{ln}) f_{ln}(\mathbf{R}_l) \right) \) where \(E_n\) is the energy consumption of node \(n\), \(e_{tx}\) and \(e_{rx}\) are energy coefficients for transmission and reception over distance \(d_{ln}\), and \(f_{ln}(\mathbf{R}_l)\) is the data flow. The analogous model's cross-domain interaction term \(\phi_{mn} (\mathbf{R}_m, \mathbf{R}_n)\), can be constructed by integrating some of the above components as follows: Interaction between domains \(m\) and \(n\) might depend on the data flows \(f_m(\mathbf{R}_m)\) and \(f_n(\mathbf{R}_n)\) over shared resources \( = \sum_{l \in \text{Shared Links}} {f_m(\mathbf{R}_m) \cdot f_n(\mathbf{R}_n)}/{C_l} \).
This expression captures the interaction effect based on the shared link capacity \(C_l\), penalizing configurations where high flows from both domains exceed link capacity. Interaction could also involve the energy consumption of nodes that are affected by multiple domains \(= \sum_{n \in \text{Shared Nodes}} \left( e_{tx}(d_{ln}) f_m(\mathbf{R}_m) \cdot f_n(\mathbf{R}_n) \right) \).
This accounts for how the combined data flows \(f_m(\mathbf{R}_m)\) and \(f_n(\mathbf{R}_n)\) over a shared node \(n\) contribute to energy consumption, influencing both domains. Lastly, the interaction can consider the impact of resource allocation on the utility functions of interconnected domains \(= \gamma_m \gamma_n \left(\mathbf{R}_m - \lambda_m\right)\left(\mathbf{R}_n - \lambda_n\right) \)
This term links the resource allocations \(\mathbf{R}_m\) and \(\mathbf{R}_n\) through their utility functions, particularly when changes in one domain's resource allocation impact the utility of another. Combining these considerations, gives the result \( \phi_{mn}(\mathbf{R}_m, \mathbf{R}_n) \). It is the graphical model's set of edges that help establish such cross connections, which results in a different optimization problem, with very likely a different optimal solution. (see Fig. \ref{fig:3})

Finally, consider the case of a large-scale virtual event in the metaverse being attended by millions of users, something which will very likely be common in the advanced metaverse. Content distribution in such a case will differ from what we are used to now. By exploiting some of the correlations among domains and using dynamic graph algorithms suggested above, the system can ensure a customized, coherent, and immersive experience for the users, which would not be possible otherwise.

\section{Discussion}

Ensuring the metaverse can handle exponentially increasing user bases and content volumes (scalability) without degradation in performance is a challenge. Managing the scalability of 3D models, environments, and interactive elements while maintaining real-time rendering capabilities is also challenging despite the advancements in graph-based algorithms. Ensuring systems can adjust in real-time to fluctuations in user activity and content creation will be tough (real-time processing). As computational intensity increases, rendering high-fidelity graphics while maintaining low latency will be challenging. Moreover, safeguarding user data from unauthorized access is essential from a security and privacy standpoint. Recent works have raised concerns regarding safety issues in the metaverse \cite{b1}, which will only worsen as it scales up. Balancing data collection for customization with user privacy and regulatory compliance involves treading a fine line. Finally, despite increased popularity, a large percentage of the world population isn't even connected to the internet, let alone such sophisticated systems. In conclusion, transitioning to an advanced metaverse presents challenges and opportunities. By employing a unified graph theoretic approach to model various aspects, we can bridge the gaps between multiple domains of the metaverse. Thus, we can better understand and optimize the current and developing metaverses by considering a holistic viewpoint instead of optimizing stand-alone systems, which will help us reach the advanced metaverse faster.


\end{document}